\documentclass[journal,onecolumn]{IEEEtran}
\usepackage{amsmath}
\usepackage{amssymb}
\usepackage{amsfonts}
\usepackage{amsxtra}
\usepackage{amstext}
\usepackage{latexsym}
\usepackage{dsfont} 
\usepackage{graphicx}

\usepackage{cite}
\DeclareMathOperator{\pr}{Pr}

\ifCLASSINFOpdf
\else
\fi

\begin{document}
%
\title{Block encryption of quantum messages}
%
%
%

\author{Min Liang and Li Yang
\thanks{Min Liang is with the Data Communication Science and Technology Research Institute, Beijing 100191, China. E-mail: liangmin07@mails.ucas.ac.cn.}
\thanks{Li Yang is with the State Key Laboratory of Information Security, Institute of Information Engineering, Chinese Academy of Sciences, Beijing 100093, China
and the University of Chinese Academy of Sciences, Beijing 100049, China. E-mail: yangli@iie.ac.cn}
}

\maketitle

\newtheorem{theorem}{Theorem}
\newtheorem{definition}{Definition}
\newtheorem{lemma}{Lemma}
\newtheorem{remark}{Remark}
\newtheorem{proposition}{Proposition}

\begin{abstract}
In modern cryptography, block encryption is a fundamental cryptographic primitive.
However, it is impossible for block encryption to achieve the same security as one-time pad.
Quantum mechanics has changed the modern cryptography, and lots of researches have shown that quantum cryptography
can outperform the limitation of traditional cryptography.

This article proposes a new constructive mode for private quantum encryption, named $\mathcal{EHE}$, which is a very simple method to construct quantum encryption from classical primitive. Based on $\mathcal{EHE}$ mode, we construct a quantum block encryption (QBE)
scheme from pseudorandom functions. If the pseudorandom functions are standard secure, our scheme is indistinguishable encryption under chosen plaintext attack. If the pseudorandom functions are permutation on the key space, our scheme can achieve perfect security. In our scheme, the key can be reused and the randomness cannot, so a $2n$-bit key can be used in an exponential number of encryptions, where the randomness will be refreshed in each time of encryption. Thus $2n$-bit key can perfectly encrypt $O(n2^n)$ qubits, and the perfect secrecy would not be broken if the $2n$-bit key is reused for only exponential times.

Comparing with quantum one-time pad (QOTP), our scheme can be the same secure as QOTP, and the secret key can be reused (no matter whether the eavesdropping exists or not). Thus, the limitation of perfectly secure encryption (Shannon's theory) is broken in the quantum setting. Moreover, our scheme can be viewed as a positive answer to the open problem in quantum cryptography ``how to unconditionally reuse or recycle the whole key of private-key quantum encryption".
In order to physically implement the QBE scheme, we only need to implement two kinds of single-qubit gates (Pauli $X$ gate and Hadamard gate), so it is within reach of current quantum technology.
\end{abstract}


%
\IEEEpeerreviewmaketitle

\section{Introduction}
%
%
%
%
\IEEEPARstart{T}{he} combination of quantum mechanics and information science forms a new science -- quantum information science, in which the information extends to quantum information. The requirement of processing quantum information occurs, and we have to develop quantum cryptographic technology for quantum information, e.g. encryption of quantum information. Since the quantum information can be seen as an extension of classical information in complex Hilbert space, the cryptographic schemes for quantum information are suitable for classical information, but not vice versa.

Quantum information encryption is a kind of basic quantum cryptographic primitive, especially the quantum one-time pad (QOTP), which has been applied in various quantum cryptographic schemes. For example, the quantum message authentication (QMA) is applied in the constructions of secure multiparty quantum computation \cite{Dupuis2012} and quantum interactive proof \cite{Aharonov2010}, and the authenticity of QMA can be guaranteed by quantum encryption \cite{Barnum2002}.

QOTP (or private quantum channel) \cite{Boykin2003,Boykin2002,Ambainis2000,Leung2001} is the first kind of quantum information encryption scheme, which uses preshared classical symmetric key and has perfect security. However, the secret key cannot be reused. The recycling issues of QOTP-key have been studied in some literatures \cite{Oppenheim2005}. Zhou et al. propose another symmetric-key encryption algorithm \cite{Zhou2006}, which uses quantum-classical hybrid keys.

Public-key encryption of quantum messages is firstly studied by Yang \cite{Yang2003}, in which both the public key and private key are classical. Because the scheme is constructed based on NP-complete problem, it has computational security at the most. Later, public-key encryption schemes with computational security are studied in more literatures \cite{Yang2010,Fujita2012,Yang2015}. In addition, public-key encryption with information-theoretic security is also studied \cite{Liang2012,Kawachi2008}.

Alagic et al.\cite{Alagic2016} propose a private-key scheme and a public-key encryption scheme for quantum data, both of which have computational security. The private-key scheme is constructed based on quantum pseudorandom function (PRF) and QOTP, but it is not indistinguishable against chosen ciphertext attack. The public-key scheme is constructed based on quantum trapdoor one-way permutation and QOTP.

There are some literatures about QMA \cite{Barnum2002,Garg2017,Portmann2017} or non-malleable quantum encryption \cite{Ambainis2009,Alagic2017}. Because authenticity of QMA implies encryption \cite{Barnum2002}, those secure quantum authentication schemes can also be used as quantum message encryption scheme; However, the secret key cannot be reused or can be recycled partially.

%
%
\subsection{Our Results}
We present a detailed description of $\mathcal{EHE}$ encryption. In the notation ``$\mathcal{EHE}$", each $\mathcal{E}$ represents a different quantum encryption operation, and $\mathcal{H}$ represents a transversal Hadamard transformation. Actually, QOTP can be viewed as a special case of $\mathcal{EHE}$ encryption, where each $\mathcal{E}$ is implemented by encrypting quantum superpositions using classical one-time pad.

\begin{figure}[!t]
\centering
\includegraphics[width=3.5in]{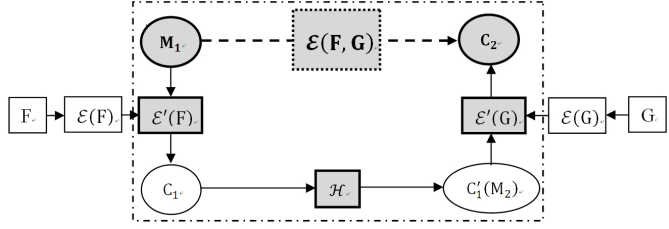}
\caption{Construction of quantum block encryption scheme $\mathcal{E}(F,G)$. The rectangles represent cryptographic primitives or related computational steps. The elliptic frames represent plaintext space or ciphertext space. The gray frames represent the detailed procedure of the scheme $\mathcal{E}(F,G)$: the quantum message in space $M_1$ is encrypted using the first scheme $\mathcal{E}'(F)$, and then be transformed using $\mathcal{H}$, and finally be encrypted using the second scheme $\mathcal{E}'(G)$.}
\label{fig1}
\end{figure}

Based on two PRFs, we construct a secure quantum block encryption (QBE) scheme in the form of $\mathcal{EHE}$ encryption. The idea is described in Fig.\ref{fig1}.
$\mathcal{E}(F)$ and $\mathcal{E}(G)$ are two classical block encryption (BE) schemes that are constructed based on two PRFs $F$ and $G$. $\mathcal{E}'(F)$ and $\mathcal{E}'(G)$ are insecure QBE schemes that are constructed using $\mathcal{E}(F)$ and $\mathcal{E}(G)$.
The whole procedure of quantum encryption $\mathcal{E}(F,G):\sigma\in M_1 \rightarrow \rho\in C_2$ can be finished in the three steps: (1) the quantum message $\sigma\in M_1$ is encrypted using the first QBE scheme $\mathcal{E}'(F)$, and the obtained ciphertext is $\rho_1\in C_1$; (2) perform transversal Hadamard transformation on $\rho_1\in C_1$, and obtain $\rho_2\in C_1'$; (3) If $C_1'\subseteq M_2$, then $\rho_2\in M_2$ can be encrypted using the second QBE scheme $\mathcal{E}'(G)$, and the obtained ciphertext is $\rho\in C_2$.

We study the security of QBE scheme $\mathcal{E}(F,G)$, and obtain the main results as follows.

\begin{theorem}[informal]\label{thm1}
If PRFs $F,G$ are chosen independently and have standard security in the quantum computation setting, then $\mathcal{E}(F,G)$ is an IND-CPA-secure QBE scheme.
\end{theorem}

\begin{theorem}[informal]\label{thm2}
$F,G$ are independent PRFs with standard security. If both $F$ and $G$ are permutations on the key space,
then $\mathcal{E}(F,G)$ is a perfectly secure QBE scheme.
\end{theorem}

Theorem \ref{thm1} states that our QBE scheme can be IND-CPA-secure. The plaintext block has the same length as ciphertext block.
Theorem \ref{thm2} states that, in some particular case, the QBE scheme can have the same security as QOTP even if the keys are reused.
Thus, our scheme can be viewed as a positive answer to an open problem in quantum cryptography ``how to unconditionally reuse or recycle the whole key of private-key quantum encryption", which has been studied in Refs.\cite{Oppenheim2005,Garg2017,Portmann2017,Damgard2005,Damgard2014,Fehr2017}.

QOTP has been widely applied in the theoretical design of various quantum encryption and authentication schemes \cite{Dupuis2012,Aharonov2010,Barnum2002,Liang2012,Portmann2017}. Based on our results, we can consider modifying those QOTP-based schemes by replacing QOTP with perfectly secure QBE, and expect an obvious optimization, for example, recycling all the keys of the scheme in Ref.\cite{Portmann2017} or lifting weak authentication to total authentication \cite{Garg2017}.

\subsection{Related works}
\subsubsection{How to construct quantum cryptographic primitives from classical ones}
Based on quantum mechanics, the information extends to quantum information, and the computation extends to quantum computation.
A natural question is whether or not the modern cryptography based on the information and computation could extend to quantum cryptography.
Concretely, how to extend classical cryptographic primitive to quantum one? Our results give an answer from the aspect of BE (or pseudorandom functions).
In addition, there are also some other related works.

In Ref.\cite{Yang2003}, a quantum public-key encryption scheme is proposed based on classical McEliece public-key cryptosystem. Later, more constructions are proposed \cite{Yang2010}. In order to improve the security, Yang and Liang \cite{Yang2015} propose the double-encryption technique, which is the origin of $\mathcal{EHE}$ encryption.

Garg et al. \cite{Garg2017} propose the ``Auth-QFT-Auth" pattern used to construct QMA scheme (denoted as $Auth_2(\mathcal{H}(Auth_1(\rho)))$), where $Auth_1,Auth_2$ are the classical Wegman-Carter MAC schemes and $\mathcal{H}$ is the quantum Hadamard transform. Obviously, this pattern is very similar to $\mathcal{EHE}$ encryption.

In fact, QOTP can be viewed as an $\mathcal{EHE}$-like construction based on classical OTP:
quantum states are encrypted using the classical one-time pad in the basis $\{|0\rangle,|1\rangle\}$,
and then using the classical one-time pad in the basis $\{|+\rangle,|-\rangle\}$.

The most related work is Ref.\cite{Alagic2016}, which propose a computationally secure framework for quantum encryption. However, their construction uses ``PRF+QOTP" mode, and our construction uses $\mathcal{EHE}$ mode. In the spirit, $\mathcal{EHE}$ mode is a special combination of two insecure encryption. This mode of combination can be extended to construct more quantum cryptographic schemes.

\subsubsection{Quantum encryption with key recycling}
OTP is a perfectly secure encryption scheme, but the key cannot be reused; In BE scheme, the key can be reused, but the security is weaker than OTP.
In quantum cryptography, there exists the same problem: QOTP has the same security as OTP, but the key cannot be reused
(Though we can use a QOTP with quantum key distribution,
this would need more rounds of interaction and more communication.). In order to settle this problem, the researchers begin to consider how to recycle part of the keys or conditionally reuse the keys.

Damgard et al.\cite{Damgard2005,Damgard2014} show how to encrypt a classical message in a quantum state and recycle the key.
Oppenheim and Horodecki \cite{Oppenheim2005} study how to encrypt a quantum message and recycle the key, and the key of QOTP can only be partially reused. Fehr and Salvail \cite{Fehr2017} propose a classical-message-oriented quantum authentication scheme with key recycling, in which the partial randomness can be extracted and be used as the OTP-key or QOTP-key. Then the combination of the authentication scheme and OTP (or QOTP) becomes a quantum encryption scheme with key recycling, and can be used to encrypt the classical or quantum information.

There are also some researches about QMA with key recycling \cite{Garg2017,Portmann2017}.
The ``Auth-QFT-Auth"  authentication scheme \cite{Garg2017} allows conditionally recycling part of the keys: the inner key can be recycled upon successful verification,
and the outer key unfortunately cannot be. Because any scheme to authenticate quantum messages
must also encrypt them \cite{Barnum2002}, these authentication schemes can also be used as encryption schemes with key recycling.

In all these schemes, the keys cannot be totally reused, and we will solve this problem through QBE scheme.

\subsection{Organization}
In Section \ref{sec:2}, we introduce some basic notations, and review three kinds of PRFs. In Section \ref{sec:23}, we describe the
$\mathcal{EHE}$ encryption technique. In Section \ref{sec:3}, we show how to construct IND-CPA-secure 
QBE scheme, and prove the perfectly secure scheme is achievable. Finally, we conclude and discuss these results.

\section{Preliminaries}
\label{sec:2}
\subsection{Notations and definitions}
$Func_n=\{f|f:\{0,1\}^n\rightarrow\{0,1\}^n\}$ denotes the set of all the functions that map $n$ bits to $n$ bits. Define $\mathcal{Y}^\mathcal{X}$ as the set of functions $\{f|f:\mathcal{X}\rightarrow\mathcal{Y}\}$, then $Func_n=\mathcal{N}^\mathcal{N}$, where $\mathcal{N}=\{0,1\}^n$.

Any classical computable function $f\in\mathcal{Y}^\mathcal{X}$ can be implemented by a quantum computer, or be implemented as an oracle which is queried on quantum superpositions.
\begin{equation}
U_f:\sum_{x\in\mathcal{X},y\in\mathcal{Y}}\alpha_{x,y}|x\rangle|y\rangle\longrightarrow\sum_{x\in\mathcal{X},y\in\mathcal{Y}}\alpha_{x,y}|x\rangle|y\oplus f(x)\rangle,
\end{equation}
where $\mathcal{X}$ and $\mathcal{Y}$ are the domain and range, respectively. $\sum_{x\in\mathcal{X},y\in\mathcal{Y}}$
can be briefly written as $\sum_{x,y}$ without leading to any misunderstanding. $\mathcal{A}^{|f\rangle}$ represents the quantum adversary $\mathcal{A}$ can access to $f$ with quantum superposition queries. $\mathcal{A}^f$ represents the (classical or quantum) adversary $\mathcal{A}$ can access to $f$ classically
\begin{equation}
O_f:(x,y)\rightarrow (x,y\oplus f(x)),\forall x\in\mathcal{X},y\in\mathcal{Y}.
\end{equation}

PRF is the basic primitive in modern cryptography. A PRF is a polynomial-time computable function $F:\mathcal{K}\times\mathcal{X}\rightarrow\mathcal{Y}$, where $\mathcal{K}$, $\mathcal{X}$ and $\mathcal{Y}$ are the key space, the domain and range, respectively. Denote $\mathcal{K}\times \mathcal{X}=\{(k,x):k\in\mathcal{K},x\in\mathcal{X}\}$. $\mathcal{K},\mathcal{X},\mathcal{Y}$ are implicit functions of the security parameter $n$. We write $y=F_k(x)$ or $y=F(k,x)$.

\begin{definition}[PRF]
A function $F:\mathcal{K}\times\mathcal{X}\rightarrow\mathcal{Y}$ is PRF, if for any probabilistic polynomial-time (PPT) adversary $\mathcal{A}$, the advantage of $\mathcal{A}$ while distinguishing between a truly random function $f$ and the function $F_k$ for a uniformly chosen $k$
\begin{equation*}
Adv_F^{PRF}(\mathcal{A})\triangleq \left|\underset{k\xleftarrow{R}\mathcal{K}}{\pr}[\mathcal{A}^{F_k}()=1]-\underset{f\xleftarrow{R} Func_n}{\pr}[\mathcal{A}^f()=1]\right|
\end{equation*}
is negligible. We write $k\xleftarrow{R}\mathcal{K}$ to represent the key $k$ is drawn from $\mathcal{K}$ uniformly and randomly. $f\xleftarrow{R} Func_n$ represents the function $f$ is uniformly chosen from $Func_n$. The notations can be briefly written as $k\leftarrow\mathcal{K}$ and $f\leftarrow Func_n$.
\end{definition}

``$\epsilon(n)$ is negligible" means that, for any polynomial $p(n)$, there exists $n_0$ such that $\epsilon(n)<\frac{1}{p(n)},\forall n>n_0$.

Pauli $X$ gate and $Z$ gate can be represented as:
$X=\left(
     \begin{array}{cc}
       0 & 1 \\
       1 & 0 \\
     \end{array}
   \right)
$,
$Z=\left(
     \begin{array}{cc}
       1 & 0 \\
       0 & -1 \\
     \end{array}
   \right)
$,
and Hadamard gate is
$H=\frac{1}{\sqrt{2}}\left(
                       \begin{array}{cc}
                         1 & 1 \\
                         1 & -1 \\
                       \end{array}
                     \right)
$.
Given any unitary matrix $U$ and a $n$-bit string $b=b_1b_2\cdots b_n$ ($b_i$ is the $i$-th bit of the string $b$), we write $U^b$ to denote $\bigotimes_{i=1}^n U^{b_i}$. Particularly, $U^{\otimes n}=\bigotimes_{i=1}^n U=U^{11\cdots 1}$.

For two $n$-bit strings $a,b\in\{0,1\}^n$, define $a\odot b=\sum_{i=1}^n a_i b_i (\mathrm{mod}2)$.

We write $[[p_k,U_k,k\in\mathcal{K}]]$ to represent a quantum message encryption scheme that performs encryption operator $U_k$ and decryption operator $U_k^\dagger$ using the symmetric key $k\in\mathcal{K}$, where $k$ is chosen with probability $p_k$ and cannot be reused. Then QOTP can be described by the notation $[[p_{ab}=\frac{1}{2^{2n}},X^a Z^b,a,b\in\{0,1\}^n]]$.

\subsection{Quantum pseudorandom functions}
Following the definitions in Ref.\cite{Zhandry2012}, there are two security notions of PRF under quantum computation model. The first notion is standard security, where the quantum adversary can only access to the function classically; We denote this kind of PRF as ``sPRF". The second one is quantum security, where the quantum adversary can access to the function with quantum superposition queries; We denote this kind of PRF as ``qPRF".

\begin{definition}[sPRF]\label{def2}
A PRF $F:\mathcal{K}\times\mathcal{X}\rightarrow\mathcal{Y}$ is standard secure, if no quantum polynomial-time (QPT) adversary
$\mathcal{A}$ making classical queries can distinguish between a truly random function $f$ and the function $F_k$ for a uniformly chosen $k$. That is,
for every such $\mathcal{A}$, there exists a negligible function $\epsilon=\epsilon(n)$ such that
$$
\left|\underset{k\leftarrow\mathcal{K}}{\pr}[\mathcal{A}^{F_k}()=1]-\underset{f\leftarrow Func_n}{\pr}[\mathcal{A}^f()=1]\right|<\epsilon.
$$
\end{definition}

\begin{definition}[qPRF]
A PRF $F:\mathcal{K}\times\mathcal{X}\rightarrow\mathcal{Y}$ is quantum secure, if no QPT adversary $\mathcal{A}$ making quantum queries can distinguish between a truly random function $f$ and the function $F_k$ for a uniformly chosen $k$. That is, for every such $\mathcal{A}$, there exists a negligible function $\epsilon=\epsilon(n)$ such that
$$
\left|\underset{k\leftarrow\mathcal{K}}{\pr}[\mathcal{A}^{|F_k\rangle}()=1]-\underset{f\leftarrow Func_n}{\pr}[\mathcal{A}^{|f\rangle}()=1]\right|<\epsilon.
$$
\end{definition}

For sPRF $F$, define $Adv_F^{sPRF}(\mathcal{A})=\left|\pr_{k\leftarrow\mathcal{K}}[\mathcal{A}^{F_k}()=1]-\pr_{f\leftarrow Func_n}[\mathcal{A}^f()=1]\right|$. For qPRF $F$, define $Adv_F^{qPRF}(\mathcal{A})=\left|\pr_{k\leftarrow\mathcal{K}}[\mathcal{A}^{|F_k\rangle}()=1]-\pr_{f\leftarrow Func_n}[\mathcal{A}^{|f\rangle}()=1]\right|$, where $\mathcal{A}$ is QPT adversary.

When quantum queries are allowed, QPT adversary has more advantage while distinguishing PRF and truly random function. That is $Adv_F^{sPRF}(\mathcal{A})<Adv_F^{qPRF}(\mathcal{A})$. If $Adv_F^{qPRF}(\mathcal{A})<\epsilon(n)$, then $Adv_F^{sPRF}(\mathcal{A})<\epsilon(n)$, where
$\epsilon(n)$ is negligible. Thus, if a PRF $F$ is a qPRF, then it is also a sPRF.

How to directly construct a sPRF that is not a qPRF? In fact, Even-Mansour block cipher is a sPRF \cite{Even1997}, but it is not a qPRF \cite{Kuwakado2012}. In addition, CBC-MAC is also not quantum-secure as a PRF \cite{Kaplan2016}.

\begin{lemma}\label{lemma1}
Given a function $G$, if $G$ is independent of PRF $\{F_k\}_{k\in\mathcal{K}}$, then
$$
\left|\underset{k\leftarrow\mathcal{K}}{\pr}[\mathcal{A}^{F_k,G}()=1]-\underset{f\leftarrow Func_n}{\pr}[\mathcal{A}^{f,G}()=1]\right|<\epsilon(n),
$$
where $\mathcal{A}$ is any PPT adversary and $\epsilon(n)$ is negligible.
\end{lemma}
\begin{IEEEproof}
Define a new quantum adversary $\mathcal{A}_G$, where the adversary $\mathcal{A}$ is allowed to access to the function $G$ classically. Because $G$ is independent of $\{F_k\}_{k\in\mathcal{K}}$, we have
\begin{equation*}
\left|\underset{k\leftarrow\mathcal{K}}{\pr}[\mathcal{A}^{F_k,G}()=1]-\underset{f\leftarrow Func_n}{\pr}[\mathcal{A}^{f,G}()=1]\right|
=\left|\underset{k\leftarrow\mathcal{K}}{\pr}[\mathcal{A}_G^{F_k}()=1]-\underset{f\leftarrow Func_n}{\pr}[\mathcal{A}_G^{f}()=1]\right|
= Adv_F^{PRF}(\mathcal{A}_G).
\end{equation*}
$F_k$ is a PRF, so $Adv_F^{PRF}(\mathcal{A}_G)$ is negligible. Thus complete the proof.
\end{IEEEproof}

There are two similar results for sPRF and qPRF, respectively.
\begin{lemma}\label{lemma2}
Given a function $G$, if $G$ is independent of sPRF $\{F_k\}_{k\in\mathcal{K}}$, then
$$
\left|\underset{k\leftarrow\mathcal{K}}{\pr}[\mathcal{A}^{F_k,G}()=1]-\underset{f\leftarrow Func_n}{\pr}[\mathcal{A}^{f,G}()=1]\right|<\epsilon(n),
$$
where $\mathcal{A}$ is any QPT adversary and $\epsilon(n)$ is negligible.
\end{lemma}

\begin{lemma}\label{lemma3}
Given a function $G$, if $G$ is independent of qPRF $\{F_k\}_{k\in\mathcal{K}}$, then
$$
\left|\underset{k\leftarrow\mathcal{K}}{\pr}[\mathcal{A}^{|F_k\rangle,|G\rangle}()=1]-\underset{f\leftarrow Func_n}{\pr}[\mathcal{A}^{|f\rangle,|G\rangle}()=1]\right|<\epsilon(n),
$$
where $\mathcal{A}$ is any QPT adversary and $\epsilon(n)$ is negligible.
\end{lemma}

\begin{remark}\label{remark1}
If $G$ is a PRF $\{G_k\}_{k\in\mathcal{K}}$ and is independent of $\{F_k\}_{k\in\mathcal{K}}$, then the results in Lemmas \ref{lemma1},\ref{lemma2} and \ref{lemma3} hold as well.
\end{remark}

\begin{theorem}[Parallel Composition]\label{thm3}
If $\{F_k\}_{k\in\mathcal{K}}$ and $\{G_k\}_{k\in\mathcal{K}}$ are two independent sPRFs, then $H_k=(F_{k_1},G_{k_2}),\forall k=k_1\parallel k_2$
is also a sPRF. That is, for any QPT adversary $\mathcal{A}$, there exists a negligible function $\epsilon(n)$ such that
$$
\left|\underset{k\leftarrow\mathcal{K}\times\mathcal{K}}{\pr}[\mathcal{A}^{H_k}()=1]-\underset{f\leftarrow Func_{2n}}{\pr}[\mathcal{A}^{f}()=1]\right|<\epsilon(n).
$$
\end{theorem}

\begin{IEEEproof}
According to Definition \ref{def2}, if $F$ is a sPRF, then for any QPT adversary $\mathcal{A}_1$ there exists a negligible function $\epsilon_1(n)$ such that
$$
\left|\underset{k\leftarrow\mathcal{K}}{\pr}[\mathcal{A}_1^{F_k}()=1]-\underset{f_1\leftarrow Func_n}{\pr}[\mathcal{A}_1^{f_1}()=1]\right|<\epsilon_1(n).
$$
If $G$ is a sPRF, then for any QPT adversary $\mathcal{A}_2$ there exists a negligible function $\epsilon_2(n)$ such that
$$
\left|\underset{k\leftarrow\mathcal{K}}{\pr}[\mathcal{A}_2^{G_k}()=1]-\underset{f_2\leftarrow Func_n}{\pr}[\mathcal{A}_2^{f_2}()=1]\right|<\epsilon_2(n).
$$
Thus for any QPT adversary $\mathcal{A}$, we have the following deduction according to Lemma \ref{lemma2} and Remark \ref{remark1}.
\begin{eqnarray*}
&&\left|\underset{k_1\leftarrow\mathcal{K},k_2\leftarrow\mathcal{K}}{\pr}[\mathcal{A}^{F_{k_1},G_{k_2}}()=1]-\underset{f_1\leftarrow Func_n,f_2\leftarrow Func_n}{\pr}[\mathcal{A}^{f_1,f_2}()=1]\right|  \\
&\leq & \left|\underset{k_1\leftarrow\mathcal{K},k_2\leftarrow\mathcal{K}}{\pr}[\mathcal{A}^{F_{k_1},G_{k_2}}()=1]-\underset{f_1\leftarrow Func_n,k_2\leftarrow\mathcal{K}}{\pr}[\mathcal{A}^{f_1,G_{k_2}}()=1]\right|  \\
&& +\left|\underset{f_1\leftarrow Func_n,k_2\leftarrow\mathcal{K}}{\pr}[\mathcal{A}^{f_1,G_{k_2}}()=1]-\underset{f_1\leftarrow Func_n,f_2\leftarrow Func_n}{\pr}[\mathcal{A}^{f_1,f_2}()=1]\right|  \\
&<& \epsilon_1(n)+\epsilon_2(n).
\end{eqnarray*}
Let $\epsilon(n)=\epsilon_1(n)+\epsilon_2(n)$, then $\epsilon(n)$ is negligible. Let $H_k=(F_{k_1},G_{k_2})$ and $f=(f_1,f_2)$. Thus complete the proof.
\end{IEEEproof}

\subsection{$\mathcal{EHE}$ encryption}\label{sec:23}
In Ref.\cite{Yang2015}, Yang and Liang have improved the security of quantum McEliece PKE using double-encryption technology. Here, the ``double-encryption" is named as ``$\mathcal{EHE}$ encryption". The new name ``$\mathcal{EHE}$ encryption" can accurately reflect its structural characteristic.

Based on $\mathcal{EHE}$ encryption, secure quantum encryption scheme can be constructed by combining two insecure ones. $\mathcal{EHE}$ is a universal technology for the construction of quantum cryptographic schemes. The basic framework can be summarized in the following three steps:
(1) Encrypt using the first insecure quantum encryption scheme; (2) Perform transversal Hadamard transformation; (3) Encrypt again using the second insecure quantum encryption scheme.

Suppose $(G_i,E_i,D_i),i=1,2$ are the two insecure quantum encryption schemes, where $G_i$,$E_i$,$D_i$ represent the key generation, encryption and decryption algorithms, respectively. $\mathcal{H}(\cdot)$ is the transversal Hadamard transformation being performed on all the input qubits.
General framework of $\mathcal{EHE}$ encryption is completely described in the following three algorithms.
\begin{itemize}
  \item $KeyGen(1^n)$: $k_1\leftarrow G_1(1^n),k_2\leftarrow G_2(1^n)$, output $k_1,k_2$;
  \item $Enc(k_1,k_2,\sigma)$: $\sigma_1\leftarrow E_1(k_1,\sigma),\sigma_2\leftarrow\mathcal{H}(\sigma_1),\rho\leftarrow E_2(k_2,\sigma_2)$, output $\rho$;
  \item $Dec(k_1,k_2,\rho)$: $\rho_1\leftarrow D_2(k_2,\rho),\rho_2\leftarrow\mathcal{H}(\rho_1),\sigma\leftarrow D_1(k_1,\rho_2)$, output $\sigma$.
\end{itemize}

The two encryption schemes $(G_i,E_i,D_i),i=1,2$ should satisfy the conditions $D_i(k_i,E_i(k_i,\sigma))=\sigma$,$\forall\sigma,i=1,2$. It is straightforward that
$$Dec(k_1,k_2,Enc(k_1,k_2,\sigma))=\sigma,\forall\sigma,$$
so the combined construction can decrypt the ciphertext correctly.

\section{Quantum block encryption}\label{sec:3}
\subsection{Some definitions}
$[[p_k,U_k,k\in\mathcal{K}]]$ is a kind of symmetric-key quantum encryption scheme, where each key $k$ is chosen with probability $p_k$ and cannot be reused. In this section, we propose the QBE scheme, which is another kind of symmetric-key scheme, and its secret key can be reused for many times.

\begin{definition}[QBE]
QBE scheme is defined by a triplet $(KeyGen,Enc,Dec)$, where $KeyGen,Enc,Dec$ are key generation, encryption and decryption algorithms, respectively.
$\mathcal{K}$ is the key space, and $\mathcal{H}_M$ and $\mathcal{H}_C$ are the quantum plaintext/ciphertext spaces. The randomness $R$ is optional.
\begin{itemize}
  \item $KeyGen$: Given a security parameter $n$, it generates a secret key $k\in\mathcal{K}$;
  \item $Enc$: Choose a random number $r\in R$ and perform the encryption transformation $Enc:\mathcal{K}\times\mathcal{H}_M\rightarrow R\times\mathcal{H}_C$ with the key $k\in\mathcal{K}$;
  \item $Dec$: Perform the decryption transformation $Dec:\mathcal{K}\times R\times\mathcal{H}_C\rightarrow\mathcal{H}_M$ with the key $k\in\mathcal{K}$.
\end{itemize}

These algorithms satisfy the condition
$Dec(k,Enc(k,\sigma))=\sigma,\forall k\in\mathcal{K},\sigma\in\mathcal{H}_M.$
\end{definition}

Similar to the security notions of classical encryption, we can define the quantum versions of indistinguishability (IND), indistinguishability against chosen plaintext attack (IND-CPA).
These definitions can also be referred to Refs.\cite{Liang2012}\cite{Alagic2016}\cite{Xiang2012}.
Notice that, indistinguishability for quantum encryption is originally defined in Ref.\cite{Xiang2012}. Later, Broadbent and Jeffery \cite{Broadbent2015} presents a definition of quantum IND-CPA with an interactive game, and gives no explicit definition of IND. Following the definition in Ref.\cite{Broadbent2015}, Ref.\cite{Alagic2016} defines IND, IND-CPA and IND-CCA with an incremental way instead of interactive game. The incremental definition is very brief and is adopted in our manuscript.

\begin{definition}[IND]\label{def5}
A QBE scheme $(KeyGen,Enc,Dec)$ is IND-secure, if for any QPT adversary $\mathcal{A}$,
\begin{equation*}
\left|Pr[\mathcal{A}(\sum_{k\in\mathcal{K}}p_k Enc(k,\sigma_1))=1]-Pr[\mathcal{A}(\sum_{k\in\mathcal{K}}p_k Enc(k,\sigma_2))=1]\right|<\epsilon(n),
\end{equation*}
where $\epsilon(n)$ is negligible, $\sigma_1,\sigma_2$ are arbitrary quantum states chosen by the adversary from $\mathcal{H}_M$, $p_k=Pr[k\leftarrow KeyGen(1^n)]$, and the probability in these terms is taken over the internal randomness of the algorithms $KeyGen$, $Enc$ and $\mathcal{A}$.
\end{definition}

Next, we introduce another definition of IND.

\begin{definition}[IND]\label{def6}
A QBE scheme $(KeyGen,Enc,Dec)$ is IND-secure, if for any QPT adversary $\mathcal{A}$,
\begin{equation*}
\left|Pr[\mathcal{A}(\sum_{k\in\mathcal{K}}p_k Enc(k,\sigma))=1]-Pr[\mathcal{A}(\sum_{k\in\mathcal{K}}p_k Enc(k,\frac{I}{2^n}))=1]\right|<\epsilon(n),
\end{equation*}
where $\epsilon(n)$ is negligible, $\sigma$ is arbitrary quantum state chosen by the adversary from $\mathcal{H}_M$,
$p_k=Pr[k\leftarrow KeyGen(1^n)]$, and the probability in these terms is taken over the internal randomness of the algorithms $KeyGen$, $Enc$ and $\mathcal{A}$.
\end{definition}

Obviously, the two definitions of IND are equivalent. The reason is as follows: (1)if a QBE scheme satisfies Definition \ref{def5}, let $\sigma_2=\frac{I}{2^n}$, then the QBE scheme satisfies Definition \ref{def6} too; (2)if a QBE scheme satisfies Definition \ref{def6}, then $|Pr[\mathcal{A}(\sum_{k\in\mathcal{K}}p_k Enc(k,\sigma_1))=1]-Pr[\mathcal{A}(\sum_{k\in\mathcal{K}}p_k Enc(k,\sigma_2))=1]|<2\epsilon(n)$ and the QBE scheme satisfies Definition \ref{def5} too.

\begin{definition}[IND-CPA]
A QBE scheme $(KeyGen,Enc,Dec)$ is IND-CPA-secure, if it is IND-secure when the QPT adversary $\mathcal{A}$ is allowed to access to the encryption oracle $Enc(k,*)$, where $k$ is the secret key.
\end{definition}

%

IND and IND-CPA define the computational security. In addition, we can define information-theoretic security, e.g. perfect security. Actually, QOTP is a kind of perfectly secure quantum encryption. In quantum cryptography, there exist some other cryptographic schemes that can achieve perfect security.

\begin{definition}[Perfect Security]
A QBE scheme $(KeyGen,Enc,Dec)$ is perfectly secure, if Definition \ref{def5} (or Definition \ref{def6}) holds for $\epsilon(n)\equiv0$ when $\mathcal{A}$ is computationally unbounded quantum adversary.
\end{definition}

In QOTP $[[p_{ab}=\frac{1}{2^{2n}},X^a Z^b,a,b\in\{0,1\}^n]]$, a secret key of
$2n$ bits is necessary for perfectly encrypting $n$ qubits. Suppose we set a restriction on $a$ and $b$ such that $a\equiv b$, then we get a new encryption scheme $[[p_c=\frac{1}{2^n},X^cZ^c,c\in\{0,1\}^n]]$. The length of the key would decrease to $n$, however, the security will also decrease.

\begin{proposition}\label{prp1}
The quantum encryption scheme $[[p_c=\frac{1}{2^n},X^c Z^c,c\in\{0,1\}^n]]$ is not IND-secure.
\end{proposition}

\begin{IEEEproof}
Suppose $n=1$. Two quantum states $\frac{1}{\sqrt{2}}(|0\rangle+i|1\rangle)$ and $|0\rangle$ are chosen as the challenge messages. Consider the two messages are encrypted. The density matrixes of the two messages are written as $\sigma_1$ and $\sigma_2$, respectively.

The key $c\in\{0,1\}$ is chosen with probability $\frac{1}{2}$. Because the adversary does not know the value of $c$, the ciphertexts corresponding to $\sigma_1$ and $\sigma_2$ should be represented as two mixed states $\rho_1$, $\rho_2$.
\begin{equation*}
\rho_1 = \sum_{c\in\{0,1\}}p_c Enc(c,\sigma_1)=\frac{1}{2}Enc(0,\sigma_1)+\frac{1}{2}Enc(1,\sigma_1) =\left(
\begin{array}{cc}
1/2 & -i/2 \\
i/2 & 1/2 \\
\end{array}
\right),
\end{equation*}
\begin{equation*}
\rho_2 = \sum_{c\in\{0,1\}}p_c Enc(c,\sigma_2)=\frac{1}{2}Enc(0,\sigma_2)+\frac{1}{2}Enc(1,\sigma_2)
= \left(
\begin{array}{cc}
1/2 & 0 \\
0 & 1/2 \\
\end{array}
\right).
\end{equation*}
The trace distance of the two ciphertexts is $D(\rho_1,\rho_2)=\frac{1}{2}$, and the adversary can efficiently distinguish the ciphertexts of $\sigma_1$ and $\sigma_2$.
In fact, the adversary chooses $\{\frac{1}{\sqrt{2}}(|0\rangle+i|1\rangle),\frac{1}{\sqrt{2}}(|0\rangle-i|1\rangle)\}$ as the measurement basis.
If the adversary measures $\rho_1$ in the basis, he can obtain $\frac{1}{\sqrt{2}}(|0\rangle+i|1\rangle)$ with probability $1$;
If the adversary measures $\rho_2$ in the basis, he can obtain $\frac{1}{\sqrt{2}}(|0\rangle+i|1\rangle)$ with probability $\frac{1}{2}$, and obtain $\frac{1}{\sqrt{2}}(|0\rangle-i|1\rangle)$ with probability $\frac{1}{2}$. Thus, the adversary can efficiently distinguish $\rho_1$ and $\rho_2$ with successful probability $\frac{3}{4}$.

For any value of $n$, we choose the two states $\frac{1}{\sqrt{2^n}}(|0\rangle+i|1\rangle)^{\otimes n}$ and $|0\rangle^{\otimes n}$ as the challenge messages, and analyze the security in the same way. Then the adversary can efficiently distinguish their ciphertexts with successful probability $1-\frac{1}{4^n}$. Thus complete the proof.
\end{IEEEproof}

\subsection{An insecure construction from classical block encryption}\label{sec:32}
Next, we introduce the PRF-based classical BE scheme $\mathcal{E}(F)$, and construct a QBE scheme $\mathcal{E}'(F)$ which is insecure.

{\bf Construction 1(Construction 5.3.9 in Ref.\cite{Goldreich2001})}: Let $F:\mathcal{K}\times\{0,1\}^n\rightarrow\{0,1\}^n$ be a PRF. Define classical BE scheme $\mathcal{E}(F)=(G_F,E_F,D_F)$ as follows.
\begin{itemize}
  \item $G_F(1^n)$: $k\xleftarrow{R}\mathcal{K}$, output $k$;
  \item $E_F(k,m)$: $r\xleftarrow{R}\{0,1\}^n,c\leftarrow m\oplus F(k,r)$, output $(r,c)$;
  \item $D_F(k,(r,c))$: $m\leftarrow c\oplus F(k,r)$, output $m$.
\end{itemize}

Based on the classical scheme $\mathcal{E}(F)$, we can construct a QBE scheme $\mathcal{E}'(F)=(G_F',E_F',D_F')$ for encrypting any $n$-qubit message.

{\bf Construction 2}: Let $\mathcal{E}(F)=(G_F,E_F,D_F)$ be a classical BE scheme defined in Construction 1, define the QBE scheme $\mathcal{E}'(F)=(G_F',E_F',D_F')$ as follows.
\begin{itemize}
  \item $G_F'(1^n)$: $k\leftarrow G_F(1^n)$, output $k$;
  \item $E_F'(k,\sigma)$: $r\xleftarrow{R}\{0,1\}^n,\rho\leftarrow X^{F(k,r)}\sigma X^{F(k,r)}$, output $(r,\rho)$;
  \item $D_F'(k,(r,\rho))$: $\sigma\leftarrow X^{F(k,r)}\rho X^{F(k,r)}$, output $\sigma$.
\end{itemize}

Assume without loss of generality that the quantum message is a pure state $\sum_m\alpha_m|m\rangle$, where $\sum_m|\alpha_m|^2=1$. According to the encryption operator $E_F'$ defined in Construction 2, the obtained ciphertext is also pure state, which can be written as $\sum_c\alpha_c|c\rangle$.
\begin{eqnarray*}
&& E_F'(k,\sum_m\alpha_m|m\rangle)=\left(r,\sum_m\alpha_m|m\oplus F(k,r)\rangle\right),\\
&& D_F'(k,(r,\sum_c\alpha_c|c\rangle))=\sum_c\alpha_c|c\oplus F(k,r)\rangle.
\end{eqnarray*}

Next we show 
that the QBE scheme $\mathcal{E}'(F)$ in Construction 2 is insecure.

\begin{theorem}\label{thm4}
The QBE scheme $\mathcal{E}'(F)=(G_F',E_F',D_F')$ in Construction 2 is not IND-secure.
\end{theorem}
\begin{IEEEproof}
Choose two quantum plaintexts $|\varphi_1\rangle=\frac{1}{\sqrt{2^n}}\sum_{m\in\{0,1\}^n}|m\rangle$ and $|\varphi_2\rangle=|0\rangle^{\otimes n}$.
Suppose the secret key is $k$, the ciphertexts of $|\varphi_1\rangle$ and $|\varphi_2\rangle$ are
\begin{eqnarray*}
E_F'(k,|\varphi_1\rangle)&=&(r,\frac{1}{\sqrt{2^n}}\sum_{m\in\{0,1\}^n}|m\oplus F(k,r)\rangle) \\
&=& (r,\frac{1}{\sqrt{2^n}}\sum_{m\in\{0,1\}^n}|m\rangle)=(r,|\varphi_1\rangle), \\
E_F'(k,|\varphi_2\rangle)&=&(r,|F(k,r)\rangle).
\end{eqnarray*}
With respect to the adversary (who does not know the key $k$), the ciphertexts of $|\varphi_1\rangle$ and $|\varphi_2\rangle$ should be written in the mixed states as follows.
\begin{eqnarray*}
\sum_{k\in\mathcal{K}}p_k E_F'(k,|\varphi_1\rangle)&=&(r,|\varphi_1\rangle\langle\varphi_1|),   \\
\sum_{k\in\mathcal{K}}p_k E_F'(k,|\varphi_2\rangle)&=&(r,\frac{1}{|\mathcal{K}|}\sum_{k\in\mathcal{K}}|F(k,r)\rangle\langle F(k,r)|).
\end{eqnarray*}
The adversary performs quantum measurement on the ciphertexts in the basis $\{|+\rangle,|-\rangle\}$. Because $|\varphi_1\rangle=|+\rangle^{\otimes n}$, while measuring its ciphertext, the outcome would be $00\cdots 0$ with probability $1$; While measuring the ciphertext of $|\varphi_2\rangle$, the outcome would be $00\cdots 0$ with probability at most $\frac{1}{|\mathcal{K}|}\sum_{k\in\mathcal{K}}\frac{1}{2^n}=\frac{1}{2^n}$. Thus, the adversary can successfully distinguish the two ciphertexts with probability at least $1-\frac{1}{2^n}$. Thus complete the proof.
\end{IEEEproof}

Theorem \ref{thm4} can be extended to the case that replacing $\mathcal{E}(F)=(G_F,E_F,D_F)$ with any quasi-length-preserving encryption scheme.
See the eprint version of Ref.\cite{Gagliardoni2016} for the definition of quasi-length-preserving encryption.

\begin{theorem}\label{thm5}
Given any quasi-length-preserving classical BE scheme, the QBE scheme constructed according to Construction 2 is not IND-secure.
\end{theorem}
\begin{IEEEproof}
The proof is similar to Theorem \ref{thm4}.
\end{IEEEproof}

From Theorems \ref{thm4} and \ref{thm5}, it is insecure to use any quasi-length-preserving classical BE schemes in the following two cases.
The first case is that the classical scheme is directly used to encrypt quantum superpositions on the quantum computer. The second case is
that the classical scheme is embedded into the quantum cryptographic protocols.

\subsection{IND-CPA quantum block encryption}\label{sec:33}
If $F$ and $G$ are PRFs, two insecure QBE schemes can be defined following the constructions in Section \ref{sec:32}. Denote the two schemes as $\mathcal{E}'(F)=(G_F',E_F',D_F')$ and $\mathcal{E}'(G)=(G_G',E_G',D_G')$, respectively. Next, we propose a secure QBE scheme $\mathcal{E}(F,G)=(KeyGen,Enc,Dec)$ following the framework of $\mathcal{EHE}$ encryption.

{\bf Construction 3}: Given two schemes $\mathcal{E}'(F)=(G_F',E_F',D_F')$ and $\mathcal{E}'(G)=(G_G',E_G',D_G')$,
define a new QBE scheme $\mathcal{E}(F,G)=(KeyGen,Enc,Dec)$ as follows.
\begin{itemize}
  \item $KeyGen(1^n)$: $k_1\leftarrow G_F'(1^n)$, $k_2\leftarrow G_G'(1^n)$, output $(k_1,k_2)$;
  \item $Enc(k_1,k_2,\sigma)$: $(r_1,\sigma_1)\leftarrow E_F'(k_1,\sigma)$,$\sigma_2\leftarrow\mathcal{H}(\sigma_1)$, $(r_2,\rho)\leftarrow E_G'(k_2,\sigma_2)$, output $(r_1,r_2,\rho)$;
  \item $Dec(k_1,k_2,(r_1,r_2,\rho))$: $\sigma_2\leftarrow D_G'(k_2,(r_2,\rho))$, $\sigma_1\leftarrow\mathcal{H}(\sigma_2)$, $\sigma\leftarrow D_F'(k_1,(r_1,\sigma_1))$, output $\sigma$.
\end{itemize}

According to the QBE scheme $\mathcal{E}(F,G)$ defined in Construction 3, we encrypt $n$ qubits $\sigma$ with the keys $k_1,k_2$, and obtain
\begin{equation}\label{eqn4}
Enc(k_1,k_2,\sigma)=(r_1,r_2,U^{F,G}_{k_1,k_2}(r_1,r_2)\sigma (U^{F,G}_{k_1,k_2}(r_1,r_2))^\dagger)\stackrel{\triangle}{=}(r_1,r_2,\rho),
\end{equation}
where $U^{F,G}_{k_1,k_2}(r_1,r_2)=X^{G(k_2,r_2)}H^{\otimes n}X^{F(k_1,r_1)}$.

We decrypt the ciphertext $(r_1,r_2,\rho)$ with the keys $k_1,k_2$, and obtain
\begin{equation}\label{eqn5}
Dec(k_1,k_2,(r_1,r_2,\rho))=(U^{F,G}_{k_1,k_2}(r_1,r_2))^\dagger\rho U^{F,G}_{k_1,k_2}(r_1,r_2).
\end{equation}
Notice that
\begin{equation}
X^{G(k_2,r_2)}H^{\otimes n}X^{F(k_1,r_1)}\sigma X^{F(k_1,r_1)}H^{\otimes n}X^{G(k_2,r_2)} =H^{\otimes n}Z^{G(k_2,r_2)}X^{F(k_1,r_1)}\sigma X^{F(k_1,r_1)}Z^{G(k_2,r_2)}H^{\otimes n}.
\end{equation}
Then we can make a slight modification to the encryption/decryption operators (in Equations (\ref{eqn4}) and (\ref{eqn5})) as follows.
\begin{eqnarray}
&& Enc(k_1,k_2,\sigma)=(r_1,r_2,Z^{G(k_2,r_2)}X^{F(k_1,r_1)}\sigma X^{F(k_1,r_1)}Z^{G(k_2,r_2)}), \label{eqn7}\\
&& Dec(k_1,k_2,(r_1,r_2,\rho))=X^{F(k_1,r_1)}Z^{G(k_2,r_2)}\rho Z^{G(k_2,r_2)}X^{F(k_1,r_1)}.    \label{eqn8}
\end{eqnarray}
It can be seen that, the only modification is that the quantum operator $H^{\otimes n}$ is discarded. Because the operator $H^{\otimes n}$ does not contain variable parameters, the modification would not affect its security essentially. However, there exists a slight disadvantage that is analyzed as follows.

Upon the modifications (defined by Equations (\ref{eqn7}) and (\ref{eqn8})), if $|m\rangle$ is encrypted with the keys $k_1,k_2$ and the randomness are $r_1,r_2$, then the ciphertext would be $|m\oplus F(k_1,r_1)\rangle$ (ignoring the global phase which depends on $G$); If the ciphertext is encrypted and the same randomness $r_1,r_2$ are used, then the original message $|m\rangle$ would be restored.
In the same way, we consider the original QBE scheme (defined by Equations (\ref{eqn4}) and (\ref{eqn5})). If $|m\rangle$ is encrypted twice in sequence using the same randomness, then we can obtain $|m\oplus F(k_1,r_1)\oplus G(k_2,r_2)\rangle$, instead of $|m\rangle$.
For this tiny difference, we decide to choose the original scheme in Construction 3. That is, the Hadamard transformation $H^{\otimes n}$ is kept in the scheme.

It can be seen that the QBE scheme $\mathcal{E}(F,G)=(KeyGen,Enc,Dec)$ is very similar to QOTP. The difference is that, the QOTP-key is replaced with the pseudorandom numbers generated from the PRFs $F,G$ with the keys $k_1,k_2$ and randomness $r_1,r_2$. According to Construction 3, the keys of the PRFs (or classical BE schemes) are used as the key of QBE scheme $\mathcal{E}(F,G)$. Because the keys of the PRFs (or classical BE schemes) can be reused, the key of $\mathcal{E}(F,G)$ can also be reused.
However, the randomness $r_1,r_2$ cannot be reused, or else the security would decrease. The proof is as follows.

\begin{proposition}\label{prp3}
For the QBE scheme $\mathcal{E}(F,G)=(KeyGen,Enc,Dec)$ defined in Construction 3, if it is allowed to reuse the randomness $(r_1,r_2)$, then the scheme is not IND-CPA-secure.
\end{proposition}
\begin{IEEEproof}
Let $k_1,k_2$ be the secret key of QBE scheme, and choose the randomness $(r_1,r_2)$. For the first time, the sender encrypts the quantum message $\sigma$, and obtains the ciphertext
\begin{equation*}
Enc(k_1,k_2,\sigma)=(r_1,r_2,U^{F,G}_{k_1,k_2}(r_1,r_2)\sigma (U^{F,G}_{k_1,k_2}(r_1,r_2))^\dagger)\stackrel{\triangle}=(r_1,r_2,\rho).
\end{equation*}
In the CPA model, the adversary is allowed to access to the quantum encryption oracle. Given the input $\rho$, the adversary can query the quantum encryption oracle $O_{Enc(k_1,k_2,*)}$. If the randomness $(r_1,r_2)$ are reused, then the adversary would obtain the new ciphertext
\begin{equation*}
O_{Enc(k_1,k_2,*)}(\rho)=(r_1,r_2,U^{F,G}_{k_1,k_2}(r_1,r_2)\rho (U^{F,G}_{k_1,k_2}(r_1,r_2))^\dagger)=(r_1,r_2,X^cZ^c\sigma Z^cX^c),
\end{equation*}
where $c\stackrel{\triangle}{=}F(k_1,r_1)\oplus G(k_2,r_2)$. The ciphertext $X^cZ^c\sigma Z^cX^c$ can be viewed as the outcome of performing quantum encryption scheme $[[p_c=\frac{1}{2^n},X^cZ^c,c\in\{0,1\}^n]]$ on the quantum message $\sigma$. From Proposition \ref{prp1}, we conclude the QBE scheme in Construction 3 is not IND-CPA-secure if the randomness is reused.
\end{IEEEproof}

According to Proposition \ref{prp3}, while applying the QBE scheme $\mathcal{E}(F,G)$, the randomness $r_1,r_2$ cannot be reused, and should be chosen randomly in every execution of encryption.

Next we prove the security of QBE scheme $\mathcal{E}(F,G)$ in Construction 3.
\begin{theorem}\label{thm6}
If $F,G:\mathcal{K}\times\{0,1\}^n\rightarrow\{0,1\}^n$ are two independent sPRFs, then $\mathcal{E}(F,G)=(KeyGen,Enc,Dec)$ in Construction 3 is an IND-CPA-secure QBE scheme.
\end{theorem}
\begin{IEEEproof}
If the scheme in Construction 3 adapts the truly random functions $f_1,f_2\in Func_n$ (instead of PRFs $F,G$), then the scheme $\mathcal{E}(f_1,f_2)$ would be the same as QOTP. So the scheme would have perfect security.

Next we show the QBE scheme is IND-secure while using the two sPRFs $F$ and $G$.

According to the QBE scheme, if totally mixed state $\frac{I}{2^n}$ is encrypted, the outcome is $(r_1,r_2,\frac{I}{2^n})$, where $r_1,r_2$ are chosen randomly.
Given any QPT adversary $\mathcal{A}$, assume $\mathcal{A}$ can distinguish the two ciphertexts of arbitrary state $\sigma$ and $\frac{I}{2^n}$ with advantage

\begin{equation}\label{eqn9}
\left|Pr\left[\mathcal{A}(r_1,r_2,\frac{1}{|\mathcal{K}|^2}\sum_{k_1,k_2}U^{F,G}_{k_1,k_2}(r_1,r_2)\sigma (U^{F,G}_{k_1,k_2}(r_1,r_2))^\dagger)=1\right]-Pr\left[\mathcal{A}(r_1,r_2,\frac{I}{2^n})=1\right]\right|=\epsilon(n).
\end{equation}

Then we prove $\epsilon(n)$ is negligible as follows. For the pair of sPRFs $(F,G)$, we construct a distinguisher $\mathcal{D}$ invoking the QPT adversary $\mathcal{A}$. The distinguisher $\mathcal{D}$ can classically query a pair of functions, and should make a judgement about the queried functions, e.g. the queried functions are a pair of PRFs $(F,G)$ or truly random functions $(f_1,f_2)$.

{\bf Construction of distinguisher $\mathcal{D}$.} $\mathcal{D}$ is given an input $1^n$ and a pair of classical random oracles $(O_1,O_2)$, where $O_i:\{0,1\}^n\rightarrow\{0,1\}^n,i=1,2$.
\begin{enumerate}
  \item Choose a pair of random values $r_1,r_2\in\{0,1\}^n$;
  \item Access to the pair of classical random oracles $(O_1,O_2)$ with input $r_1,r_2$, and obtain the outcome $(s_1,s_2)=(O_1(r_1),O_2(r_2))$;
  \item Randomly choose a plaintext $\sigma$ ($\sigma\neq\frac{I}{2^n}$). The output $(s_1,s_2)$ is used as the key to encrypt $\sigma$ as follow: $\sigma\rightarrow(r_1,r_2,X^{s_2}H^{\otimes n}X^{s_1}\sigma X^{s_1}H^{\otimes n}X^{s_2})$; Denote the ciphertext as $(r_1,r_2,\rho)$;
  \item Invoke the QPT adversary $\mathcal{A}$ on input $(r_1,r_2,\rho)$, and output whatever $\mathcal{A}$ does.
\end{enumerate}
In the above distinguisher, $\mathcal{D}$ may access two kinds of classical random oracles. The first one is for truly random functions $(f_1,f_2)$, and the second one is for PRFs $(F,G)$. We discuss the two cases as follows.
\begin{description}
  \item[(a)] If $\mathcal{D}$ access to the truly random functions $(f_1,f_2)$, then $(s_1,s_2)$ is a random element in $\{0,1\}^{2n}$. In addition, the value of $(s_1,s_2)$ is not accessible to $\mathcal{A}$ in the distinguisher. From the aspect of $\mathcal{A}$, the ciphertext $(r_1,r_2,\rho)$ can be written as a mixed state $(r_1,r_2,\frac{1}{2^{2n}}\sum_{s_1,s_2}X^{s_2}H^{\otimes n}X^{s_1}\sigma X^{s_1}H^{\otimes n}X^{s_2})$ (That is $(r_1,r_2,\frac{I}{2^n})$). Thus,
      \begin{equation}\label{eqn10}
      Pr[\mathcal{D}^{f_1,f_2}()=1]=Pr[\mathcal{A}(r_1,r_2,\frac{I}{2^n})=1],
      \end{equation}
      where $f_1,f_2$ are chosen randomly and independently from the set $Func_n$.
  \item[(b)] If $\mathcal{D}$ access to PRFs $(F,G)$, then $(s_1,s_2)=(F(k_1,r_1),G(k_2,r_2))$. From the aspect of $\mathcal{A}$ (who does not know $k_1,k_2$), the ciphertext $(r_1,r_2,\rho)$ can be written as
      $(r_1,r_2,\frac{1}{|\mathcal{K}|^2}\sum_{k_1,k_2}U^{F,G}_{k_1,k_2}(r_1,r_2)\sigma (U^{F,G}_{k_1,k_2}(r_1,r_2))^\dagger)$. It can be concluded that
      \begin{equation}\label{eqn11}
      Pr[\mathcal{D}^{F_{k_1},G_{k_2}}()=1]=Pr[\mathcal{A}(r_1,r_2,\frac{1}{|\mathcal{K}|^2}\sum_{k_1,k_2}U^{F,G}_{k_1,k_2}(r_1,r_2)\sigma (U^{F,G}_{k_1,k_2}(r_1,r_2))^\dagger)=1],
      \end{equation}
      where $k_1,k_2\in\mathcal{K}$ are chosen randomly and independently.
\end{description}
From the equations (\ref{eqn9})(\ref{eqn10})(\ref{eqn11}), it can be deduced that
\begin{equation}\label{eqn12}
|Pr[\mathcal{D}^{F_{k_1},G_{k_2}}()=1]-Pr[\mathcal{D}^{f_1,f_2}()=1]|=\epsilon(n).
\end{equation}
$\mathcal{A}$ is a QPT algorithm, then the distinguisher $\mathcal{D}$ invoking $\mathcal{A}$ is also a QPT algorithm. Using Theorem \ref{thm3}, if $F,G$ are sPRFs,
then $\epsilon(n)$ in Equation (\ref{eqn12}) is negligible. From Equation (\ref{eqn9}) and Definition \ref{def6}, the QBE scheme $\mathcal{E}(F,G)$ is IND-secure.

Consider the case that the adversary $\mathcal{A}$ is allowed to access to quantum encryption oracle $$O_{Enc(k_1,k_2,\ast)}:\sigma\rightarrow(r_1,r_2,U^{F,G}_{k_1,k_2}(r_1,r_2)\sigma (U^{F,G}_{k_1,k_2}(r_1,r_2))^\dagger).$$ If the randomness used by $O_{Enc(k_1,k_2,\ast)}$ have also been used in challenge query, then it would be insecure (According to Proposition \ref{prp3}, the advantage of $\mathcal{A}$ while distinguishing the challenge ciphertexts would be non-negligible).
However, the encryption oracle will use a fresh randomness that is chosen uniformly and independently, so the probability that $O_{Enc(k_1,k_2,\ast)}$ uses the same randomness as the challenge query is negligible.
Then allowing $\mathcal{A}$ to access to encryption oracle $O_{Enc(k_1,k_2,\ast)}$ has negligible effect on all the above proof of IND security. Thus the QBE scheme $\mathcal{E}(F,G)$ is IND-CPA-secure.
\end{IEEEproof}

\begin{remark}
From the proof of Theorem \ref{thm6}, the distinguisher can classically access to the oracles of PRFs (or truly random functions). The PRFs are not required to have quantum security. The PRFs with standard security are sufficient to assure the IND security of the QBE scheme.
\end{remark}

Corollary 3.6.7 in Ref.\cite{Goldreich2001} has shown that the existence of one-way function implies the existence of PRF. Zhandry \cite{Zhandry2012} has proved that, if PRF exists then there exists sPRF that is not qPRF. Thus, from Theorem \ref{thm6}, we reduce IND-CPA-secure QBE scheme to the existence of one-way function. That is, if there exist one-way functions, then IND-CPA-secure QBE schemes exist as well.

\begin{definition}\label{def10}
A function $F:\mathcal{K}\times\{0,1\}^n\rightarrow\{0,1\}^n$ is pairwise independent sPRF, if the two probability distributions $(F_{k_1}(U_n),F_{k_2}(U_n))$,$k_1,k_2\in\mathcal{K}$ and $f(U_{2n})$  are QPT-indistinguishable, where $U_n$ is uniformly distributed over $\{0,1\}^n$ and $f$ is a truly random function in $Func_{2n}$. That is
\begin{equation*}
\left|\underset{(k_1,k_2)\leftarrow\mathcal{K}\times\mathcal{K}}{\pr}[\mathcal{A}^{F_{k_1},F_{k_2}}()=1]-\underset{f\leftarrow Func_{2n}}{\pr}[\mathcal{A}^f()=1]\right|<\epsilon(n),
\end{equation*}
where $\epsilon(n)$ is negligible, and $\mathcal{A}$ is any QPT adversary. $\mathcal{A}$ accesses to the two functions $F_{k_1}(*),F_{k_2}(*)$ with two independent inputs (the two inputs may be the same or different).
\end{definition}

If $F$ is a pairwise independent PRF, let $G=F$, then a QBE scheme $\mathcal{E}(F,F)=(KeyGen,Enc,Dec)$ can be constructed from $\mathcal{EHE}$ encryption technology.

{\bf Construction 4}: Given a pairwise independent PRF $F:\mathcal{K}\times\{0,1\}^n\rightarrow\{0,1\}^n$, an insecure QBE scheme $\mathcal{E}'(F)=(G_F',E_F',D_F')$ can be constructed following Constructions 1 and 2. Then a secure QBE scheme $\mathcal{E}(F,F)=(KeyGen,Enc,Dec)$ can be constructed as follows.
\begin{itemize}
  \item $KeyGen(1^n)$: $k_1\leftarrow G_F'(1^n)$, $k_2\leftarrow G_F'(1^n)$, output $(k_1,k_2)$;
  \item $Enc(k_1,k_2,\sigma)$: $(r_1,\sigma_1)\leftarrow E_F'(k_1,\sigma)$, $\sigma_2\leftarrow\mathcal{H}(\sigma_1)$, $(r_2,\rho)\leftarrow E_F'(k_2,\sigma_2)$, output $(r_1,r_2,\rho)$;
  \item $Dec(k_1,k_2,(r_1,r_2,\rho))$: $\sigma_2\leftarrow D_F'(k_2,(r_2,\rho))$, $\sigma_1\leftarrow\mathcal{H}(\sigma_2)$, $\sigma\leftarrow D_F'(k_1,(r_1,\sigma_1))$, output $\sigma$.
\end{itemize}

\begin{theorem}
    If $F$ is a pairwise independent PRF and has standard security, then $\mathcal{E}(F,F)$ in Construction 4 is an IND-CPA-secure QBE scheme.
\end{theorem}
\begin{IEEEproof}
    The proof is similar to Theorem \ref{thm6}. Definition \ref{def10} is used in the proof. The details are omitted.
\end{IEEEproof}

\subsection{Perfectly secure case}\label{sec:35}
In Section \ref{sec:33}, the QBE scheme in Construction 3 has been proved to be IND-CPA-secure. Next we show the QBE scheme can achieve higher security in a particular case.

It is well known that, BE cannot achieve the same security as OTP in classical cryptography. However, based on quantum mechanics, there may be an important breakthrough -- QBE can achieve the same security as QOTP. Next we show the QBE scheme $\mathcal{E}(F,G)=(KeyGen,Enc,Dec)$ can achieve perfect security in certain special case.

\begin{theorem}\label{thm10}
Given two independent sPRFs $F,G:\mathcal{K}\times \mathcal{X}\rightarrow \mathcal{Y}$, where $\mathcal{K}=\mathcal{X}=\mathcal{Y}=\{0,1\}^n$, if for any fixed $x$, both $F(*,x):\mathcal{K}\rightarrow \mathcal{Y}$ and $G(*,x):\mathcal{K}\rightarrow \mathcal{Y}$ are permutations, then $\mathcal{E}(F,G)$ in Construction 3 is a perfectly secure QBE scheme.
\end{theorem}

Notice that, Theorem 10 proves a special case of the scheme in Theorem 6 with only one additional limitation on the functions $F,G$. So the reusability of the key would not be affected. We have presented a strict proof that, the security is enhanced with this additional limitation, and achieve the same level as QOTP.

\begin{IEEEproof}
    From Theorem \ref{thm6}, $\mathcal{E}(F,G)$ in Construction 3 is an IND-CPA-secure QBE scheme. Next we prove it can achieve perfect security if $F(*,x)$ and $G(*,x)$ are permutations.

    Suppose a block of quantum plaintext has $n$ qubits, and its density operator $\sigma$ can be written as a $2^n\times 2^n$ matrix with trace $tr(\sigma)=1$. Given a set of all $2^n\times 2^n$ matrixes, it is an inner space if we define inner product as $(M_1,M_2)=tr(M_1 M_2^\dagger)$, where $M_1$ and $M_2$ are $2^n\times 2^n$ matrixes. Then the set $\{X^\alpha Z^\beta |\alpha,\beta\in\{0,1\}^n \}$ is a group of complete orthogonal bases. Thus the density operator $\sigma$ can be expressed as $\sigma=\sum_{\alpha,\beta}a_{\alpha,\beta}X^\alpha Z^\beta$, where $a_{\alpha,\beta}=\frac{1}{2^n}tr(\sigma Z^\beta X^\alpha)$. According to the QBE scheme $\mathcal{E}(F,G)$, quantum plaintext $\sigma$ is encrypted with the keys $k_1,k_2\in\{0,1\}^n$ as follows.
    \begin{equation*}
    Enc(k_1,k_2,\sigma)=(r_1,r_2,\sum_{\alpha,\beta}a_{\alpha,\beta}U^{F,G}_{k_1,k_2}(r_1,r_2)X^\alpha Z^\beta (U^{F,G}_{k_1,k_2}(r_1,r_2))^\dagger).
    \end{equation*}

    The keys $k_1,k_2$ are unknown by the adversary and every $k_1,k_2$ are used with identical probability. Thus, from the aspect of the adversary, the quantum ciphertext should be represented as an equal mixture of a quantum plaintext $\sigma$ encrypted under all possible keys with uniform probability
    \begin{equation*}
    \frac{1}{2^{2n}}\sum_{k_1,k_2}Enc(k_1,k_2,\sigma)=(r_1,r_2,\frac{1}{2^{2n}}\sum_{\alpha,\beta}a_{\alpha,\beta}\sum_{k_1,k_2}U^{F,G}_{k_1,k_2}(r_1,r_2)X^\alpha Z^\beta (U^{F,G}_{k_1,k_2}(r_1,r_2))^\dagger).
    \end{equation*}
    Using the following three equations
    \begin{eqnarray}
    Z^\beta X^{F(k_1,r_1)} &=& (-1)^{\beta\odot F(k_1,r_1)}X^{F(k_1,r_1)}Z^\beta,  \\
    H^{\otimes n}X^\alpha Z^\beta H^{\otimes n} &=& Z^\alpha X^\beta,  \\
    X^{G(k_2,r_2)}Z^\alpha &=& (-1)^{\alpha\odot G(k_2,r_2)}Z^\alpha X^{G(k_2,r_2)},
    \end{eqnarray}
    one can conclude that
    \begin{equation}
    \frac{1}{2^{2n}}\sum_{k_1,k_2}Enc(k_1,k_2,\sigma)=(r_1,r_2,\frac{1}{2^{2n}}\sum_{\alpha,\beta}a_{\alpha,\beta}\sum_{k_1,k_2}(-1)^{\beta\odot F(k_1,r_1)}(-1)^{\alpha\odot G(k_2,r_2)} Z^\alpha X^\beta ).
    \end{equation}
    If $F(*,r_1):\mathcal{K}\rightarrow\mathcal{Y}$ and $G(*,r_2):\mathcal{K}\rightarrow\mathcal{Y}$ are permutations, then
    \begin{eqnarray}
    \frac{1}{2^n}\sum_{k_1}(-1)^{\beta\odot F(k_1,r_1)} &=& \delta_{\beta,0},\forall\beta\in\{0,1\}^n, \label{eqn20}\\
    \frac{1}{2^n}\sum_{k_2}(-1)^{\alpha\odot G(k_2,r_2)} &=& \delta_{\alpha,0},\forall\alpha\in\{0,1\}^n,  \label{eqn21}
    \end{eqnarray}
    where the function $\delta_{x,y}=\left\{
                            \begin{array}{ll}
                              1, & \hbox{x=y;} \\
                              0, & \hbox{otherwise.}
                            \end{array}
                          \right. $
    Using Equations (\ref{eqn20})(\ref{eqn21}), it can be deduced that
    \begin{equation}\label{eqn22}
    \frac{1}{2^{2n}}\sum_{k_1,k_2}Enc(k_1,k_2,\sigma)=(r_1,r_2,\sum_{\alpha,\beta}a_{\alpha,\beta}\delta_{\alpha,0}\delta_{\beta,0}Z^\alpha X^\beta) =(r_1,r_2,a_{0,0}I)=(r_1,r_2,\frac{tr(\sigma)}{2^n}I)=(r_1,r_2,\frac{I}{2^n}).
    \end{equation}
    The facts $a_{\alpha,\beta}=tr(\sigma Z^\beta X^\alpha)/2^n$ and $tr(\sigma)=1$ are used in the above deduction. $r_1,r_2$ are randomly chosen and are independent of the plaintext. Then the adversary can obtain nothing from the quantum ciphertext $(r_1,r_2,\frac{I}{2^n})$. Thus the QBE scheme $\mathcal{E}(F,G)$ has perfect security.
\end{IEEEproof}

Because the perfectly secure QBE scheme is just a special case of the constructions in previous sections, the related results and discussions in Section \ref{sec:33}
are also suitable for the perfectly secure QBE scheme. So the keys $k_1,k_2$ are reusable and would not decrease the security. If the randomness $(r_1,r_2)$ are reused, the security would decrease.

Notice that the key can be reused and the randomness cannot. The randomness $r_1,r_2$ has exponential different choices, so a $2n$-bit key can be used in an exponential number of encryptions, where the randomness will be refreshed in each time of encryption. Thus $2n$-bit key can perfectly encrypt $O(n2^n)$ qubits, and the perfect secrecy would not be broken if the $2n$-bit key is reused only exponential times.

\begin{remark}
In Theorem \ref{thm10}, the functions $F,G$ should satisfy two conditions: (1) they are independent sPRFs; (2) for any fixed $x$, both $F(*,x)$ and $G(*,x)$ are permutations. We argue that $\mathcal{E}(F,G)$ cannot be a perfectly secure QBE if the condition (1) does not hold. For example, let $F(k_1,r_1)=k_1\oplus r_1$ and $G(k_2,r_2)=k_2\oplus r_2$, then both $F(*,r_1)$ and $G(*,r_2)$ are permutations. So $Enc(k_1,k_2,\sigma)=(r_1,r_2,H^{\otimes n}Z^{k_2\oplus r_2}X^{k_1\oplus r_1}\sigma X^{k_1\oplus r_1}Z^{k_2\oplus r_2}H^{\otimes n})$.
Because $r_1,r_2$ are public, the encryption is equivalent to $QOTP(k_1,k_2,\sigma)=Z^{k_2}X^{k_1}\sigma X^{k_1}Z^{k_2}$. Thus the keys $k_1,k_2$ cannot be reused, and $\mathcal{E}(F,G)$ is not a QBE.
\end{remark}

\begin{remark}
In Theorem \ref{thm10}, the sPRF $F$ (or $G$) is required that, for any fixed $x$, the function $F(*,x)$ (or $G(*,x)$) is permutation on the key space $\mathcal{K}=\{0,1\}^n$. In fact, a good candidate is GGM-PRF \cite{Goldreich1986} construction $\mathrm{PRF}_k(x_1x_2\cdots x_n)=G_{x_1}(\cdots G_{x_{n-1}}(G_{x_n}(k))\cdots), x_i\in\{0,1\}$, where $G_0(x),G_1(x)$ are pseudorandom permutation from $\mathcal{K}$ to $\mathcal{K}$.
\end{remark}

Next, we give a detailed comparison between our scheme and QOTP, especially their relations and differences. (1) For QOTP (see Ref.\cite{Boykin2003}), while considering the encryption of $n$ qubits, we should use an unused $2n$-bit key in each encryption, and an used key may be chosen again with probability $\frac{1}{2^{2n}}$ if the key is randomly chosen. For our scheme, the key can be reused, but a $2n$-bit randomness should be sampled and an used randomness may be chosen again with probability $\frac{1}{2^{2n}}$. (2) In QOTP, the key can be used only one time and no randomness is used. In our scheme, the key can be reused, and we only need to choose a $2n$-bit randomness in each encryption. Because the randomness can be chosen from exponential candidates, our scheme can be viewed as exponential times of $n$-qubit QOTP encryption with the same key. (3) In the $n$-qubit QOTP, the key has $2n$ bits, where $n$ can be arbitrary value. That means the length of the key is variable. In our scheme, the randomness has $2n$ bits, where the value $n$ depends on the length of the key. (4) In QOTP, $2n$-bit key can perfectly encrypt $n$ qubits. In our scheme, $2n$-bit key can perfectly encrypt $O(n2^n)$ qubits, since the scheme would not be perfectly secure when the randomness is reused. (5) Our scheme can be implemented using Pauli $X$ and $H$ gates, and the number is at most $3n$ ($n$ is the length of one block); the QOTP can be implemented using Pauli $X$ gate and $Z$ gate, and the number is at most $2n$. Thus, the QBE scheme has nearly the same difficulty and complexity as QOTP from the aspect of physical implementation. (6) QOTP can be completely replaced with our scheme. Currently, QOTP has been used as a basic quantum primitive in various cryptographic protocols and algorithms \cite{Dupuis2012,Aharonov2010,Barnum2002,Liang2012,Portmann2017}. If the QOTP in these protocols or algorithms is replaced with perfectly secure QBE scheme, then optimized schemes could be obtained.

\subsection{Multiple-message encryption}
Given a classical BE scheme, if it is IND-CPA-secure in single-message encryption, then it is also IND-CPA-secure in multiple-message encryption. However, it is not the case for QBE scheme: QBE scheme is secure in single-message encryption, however it may be insecure in multiple-message encryption.

In this section, we show that the perfectly secure QBE scheme $\mathcal{E}(F,G)$ is not perfectly secure in multiple-message encryption. However, the multiple-message encryption would be perfectly secure if the QBE scheme is applied in the operation mode ``encrypt-decrypt-confirm": Alice encrypts one block of message and sends it to Bob; After receiving Alice's one block of ciphertext, Bob decrypts it; Next, Bob confirms publicly that he has decrypted the ciphertext; Then Alice and Bob start the next round of encrypted communication.

Given $s(n)$ blocks of messages $\bigotimes_{1\leq i\leq s(n)}\sigma_i$, from the aspect of the adversary, the multiple-message encryption would output
\begin{equation}\label{eqn31}
    \left(\{(r_1^i,r_2^i)\}_{i=1}^{s(n)},\frac{1}{2^{2n}}\sum_{k_1,k_2}\bigotimes_{i=1}^{s(n)}U_{k_1,k_2}^{F,G}(r_1^i,r_2^i)\sigma_i U_{k_1,k_2}^{F,G}(r_1^i,r_2^i)^\dagger\right),
\end{equation}
where the randomness $(r_1^i,r_2^i)$ is used in the encryption of $i$th block of messages.

Generally, Equation (\ref{eqn31}) does not equal to the following equation
\begin{equation}\label{eqn32}
    \left(\{(r_1^i,r_2^i)\}_{i=1}^{s(n)},\bigotimes_{i=1}^{s(n)}\frac{1}{2^{2n}}\sum_{k_1,k_2}U_{k_1,k_2}^{F,G}(r_1^i,r_2^i)\sigma_i U_{k_1,k_2}^{F,G}(r_1^i,r_2^i)^\dagger\right),
\end{equation}

The result in the proof of Theorem \ref{thm10} does not hold any more. Thus, from the aspect of the adversary, the ciphertext of $s(n)$ blocks of messages are not equal to
\begin{equation*}
\left(\{(r_1^i,r_2^i)\}_{i=1}^{s(n)},\underbrace{\frac{I_{2^n}}{2^n}\otimes \frac{I_{2^n}}{2^n}\otimes \cdots \otimes \frac{I_{2^n}}{2^n}}_{s(n)}\right)
\end{equation*}

Thus, the perfectly secure QBE scheme $\mathcal{E}(F,G)$ is not perfectly secure in multiple-message encryption. If it works in the operation mode ``encrypt-decrypt-confirm", then the encryption of the different blocks would be independent, and it would still be perfectly secure in every single-message encryption.

\section{Conclusions and discussions}
The $\mathcal{EHE}$ encryption has been described and be used in the construction of QBE scheme. Firstly, we show how to construct an insecure QBE scheme based on PRF. Then, we propose a secure construction from two insecure QBE schemes according to $\mathcal{EHE}$  encryption. It is shown that the QBE scheme is IND-CPA-secure if there exist PRFs with standard security. 
Finally, we show the QBE scheme can have the same security as QOTP when the PRFs satisfy an additional condition.

As is well known that, ``QKD+OTP" can perfectly encrypt classical messages in theory, and there are many applications in practice. However, lots of interaction and communication are necessary, and the efficiency would decrease. Actually, the QBE scheme can also be used to encrypt classical messages. For example, the classical message $m$ can be viewed as a quantum state $|m\rangle$, and each bit $m_i$ is encrypted to a qubit $X^{G(k_2,r_2)_i}HX^{F(k_1,r_1)_i}|m_i\rangle$, which belongs to the set $\{|0\rangle,|1\rangle,|+\rangle,|-\rangle\}$. Then, while encrypting classical messages, we can use a perfectly secure QBE scheme. Because no interaction is needed in QBE scheme, it would be more efficient than ``QKD+OTP", and is a potential replacement of ``QKD+OTP" in the future. Theoretically, $2n$-bit key can perfectly encrypt $O(n2^n)$ classical bits.

For perfect secrecy, Ref.\cite{Shannon1949} proposed a strict mathematical proof that the key must have at least the same length as the plaintext. In Section \ref{sec:35}, we have shown the BE scheme based on quantum mechanics can break the limitation of perfectly secure encryption. In QOTP, $2n$-bit key
is necessary to perfectly encrypt $n$ qubits. However, in the QBE scheme, $2n$-bit key can be reused and the fresh randomness $(r_1,r_2)$ are used to encrypt another $n$ qubits, thus $2n$-bit key can be used to perfectly encrypt $O(n2^n)$ qubits.

$\mathcal{EHE}$ encryption is a kind of generic transformation used for the construction of quantum encryption scheme. It can convert classical encryption or insecure quantum encryption scheme into secure quantum encryption scheme. The QBE scheme constructed based on $\mathcal{EHE}$ encryption can be seen as an extension of classical BE scheme, and it is also suitable for encryption of the classical messages. Thus, $\mathcal{EHE}$ encryption has established the direct connection between the quantum and classical BE schemes.

Finally, two problems are left for the future research.
\begin{itemize}
  \item Construct more cryptographic schemes in the $\mathcal{EHE}$-like way. It is proved that Wegman-Carter MAC is insecure while authenticating quantum message $Auth(\rho)$ \cite{Boneh2013}, however, it can be converted into a secure QMA scheme in the $Auth_2(H(Auth_1(\rho)))$ pattern \cite{Garg2017}. In addition, our results show that $\mathcal{EHE}$  encryption can convert an insecure QBE scheme into a secure QBE scheme. Is there any other quantum cryptographic scheme that can be constructed in the $\mathcal{EHE}$-like way?
  \item Replace the QOTP with the QBE in those QOTP-based (encryption, authentication or others) schemes. QOTP has been used as an important building block in many quantum schemes. Because the perfectly secure QBE scheme in Section \ref{sec:35} has many advantages, we could replace the QOTP with the QBE and expect an obvious optimization, for example, recycling all the keys of the scheme in Ref.\cite{Portmann2017} or lifting weak authentication to total authentication \cite{Garg2017}.
\end{itemize}

\section*{Acknowledgment}
This work was supported by the National Natural Science Foundation of China (Grant No. 61672517), and National Cryptography Development Fund (Grant No. MMJJ20170108).

\ifCLASSOPTIONcaptionsoff
  \newpage
\fi


\begin{thebibliography}{1}

\bibitem{Dupuis2012}Dupuis, F., Nielsen, J.B., Salvail, L.: Actively secure two-party evaluation of any quantum operation. In: Safavi-Naini, R., Canetti, R. (eds.) CRYPTO 2012. LNCS, vol. 7417, pp. 794每811. Springer, Heidelberg (2012)
\bibitem{Aharonov2010}Aharonov, D., Ben-Or, M., Eban, E.: Interactive proofs for quantum computations. In: Proceedings of Innovations in Computer Science, ICS 2010, pp. 453每469. Tsinghua University Press (2010)
\bibitem{Barnum2002}Barnum, H., Crepeau, C., Gottesman, D., Smith, A., Tapp, A.: Authentication of quantum messages. In: Proceedings of the 43rd Symposium on Foundations of Computer Science, FOCS 2002, pp. 449每458. IEEE (2002)
\bibitem{Boykin2003}Boykin, P., Roychowdhury, V.: Optimal Encryption of Quantum Bits. Phys. Rev. A 67(4), 42317 (2003)
\bibitem{Boykin2002}Boykin, P.: Information security and quantum mechanics: security of quantum protocols. Dissertation for the Doctoral Degree. University of California, Los Angeles (2002)
\bibitem{Ambainis2000}Ambainis, A., Mosca, M., Tapp, A., De Wolf, R.: Private quantum channels. In: 41st IEEE FOCS, pp. 547-553 (2000)
\bibitem{Leung2001}Leung, D.: Quantum Vernam cipher. Quantum Inf. Comput. 2(1), 14每34 (2002)
\bibitem{Oppenheim2005}Oppenheim, J., Horodecki, M.: How to reuse a one-time pad and other notes on authentication, encryption, and protection of quantum information. Phys. Rev. A 72, 042309 (2005)
\bibitem{Zhou2006}Zhou, N. R., Liu, Y., Zeng, G. H., Xiong, J., Zhu, F. C.: Novel qubit block encryption algorithm with hybrid keys. Physica A 375(2), 693 - 698 (2006)
\bibitem{Yang2003}Yang, L.: Quantum public-key cryptosystem based on classical NP-complete problem. Manuscript (2003). arXiv: quant-ph/0310076
\bibitem{Yang2010}Yang, L., Liang, M., Li, B., Hu, L., Feng, D. G.: Quantum public-key cryptosystems based on induced trapdoor one-way transformations. Manuscript (2010). arXiv:1012.5249v2
\bibitem{Fujita2012}Fujita, H.: Quantum McEliece public-key cryptosystem. Quantum Inf. Comput. 12(3\&4), 181-202 (2012)
\bibitem{Yang2015}Yang, L., Liang, M.: Quantum McEliece public-key encryption scheme. Manuscript (2015). arXiv:1501.04895v1
\bibitem{Liang2012}Liang, M., Yang, L.: Public-key encryption and authentication of quantum information. Sci. China-Phys. Mech. Astron. 55, 1618-1629 (2012)
\bibitem{Kawachi2008}Kawachi, A., Portmann, C.: On the power of quantum encryption keys. In: J. Buchmann and J. Ding (Eds.) PQCrypto 2008. LNCS, vol. 5299, pp. 165每180 (2008)
\bibitem{Alagic2016}Alagic, G., Broadbent, A., Fefferman, B., Gagliardoni, T., Schaffner, C., Jules, M. St.: Computational Security of Quantum Encryption. In: A.C.A. Nascimento and P. Barreto (Eds.) ICITS 2016. LNCS, vol. 10015, pp. 47-71 (2016)
\bibitem{Garg2017}Garg, S., Yuen, H., Zhandry, M.: New security notions and feasibility results for authentication of quantum data. In: J. Katz and H. Shacham (Eds.) CRYPTO 2017 Part II. LNCS, vol. 10402, pp. 342每371 (2017)
\bibitem{Portmann2017}Portmann, C.: Quantum authentication with key recycling. In: J.-S. Coron and J.B. Nielsen (Eds.) EUROCRYPT 2017 Part III. LNCS, vol. 10212, pp. 339每368 (2017)
\bibitem{Ambainis2009}Ambainis, A., Bouda, J., Winter, A.: Nonmalleable encryption of quantum information. J. Math. Phys. 50(4), 042106 (2009)
\bibitem{Alagic2017}Alagic, G., Majenz, C.: Quantum non-malleability and authentication. In: J. Katz and H. Shacham (Eds.) CRYPTO 2017 Part II. LNCS, vol. 10402, pp. 310每341 (2017)
\bibitem{Damgard2005}Damgard, I., Pedersen, T.B., Salvail, L.: A quantum cipher with near optimal key-recycling. In: Shoup, V. (ed.) CRYPTO 2005. LNCS, vol. 3621, pp. 494-510. Springer, Heidelberg (2005)
\bibitem{Damgard2014}Damgard, I., Brochmann Pedersen, T., Salvail, L.: How to re-use a one-time pad safely and almost optimally even if P=NP. Nat. Comput. 13(4), 469-486 (2014)
\bibitem{Fehr2017}Fehr, S., Salvail, L.: Quantum authentication and encryption with key recycling. In: J.-S. Coron and J.B. Nielsen (Eds.) EUROCRYPT 2017 Part III. LNCS, vol. 10212, pp. 311-338 (2017)
\bibitem{Zhandry2012}Zhandry, M.: How to Construct Quantum Random Functions. In: 53rd IEEE FOCS, pp. 679-687 (2012)
\bibitem{Even1997}Even, S., Mansour, Y.: A construction of a cipher from a single pseudorandom permutation. J. Cryptology 10(3), 151-162 (1997)
\bibitem{Kuwakado2012}Kuwakado, H., Morii, M.: Security on the quantum-type Even-Mansour cipher. In: Proceedings of the International Symposium on Information Theory and Its Applications (ISITA), pp. 312每316. IEEE Computer Society (2012)
\bibitem{Kaplan2016}Kaplan, M., Leurent, G., Leverrier, A., Naya-Plasencia, M.: Breaking symmetric cryptosystems using quantum period finding. In: Robshaw, M., Katz, J. (eds.) CRYPTO 2016. LNCS, vol. 9815, pp. 207-237. Springer, Heidelberg (2016).
\bibitem{Xiang2012}Xiang, C., Yang, L.: Indistinguishability, semantic security for quantum encryption scheme. In: Proceedings of SPIE, vol. 8554, p.85540G-8 (2012)
\bibitem{Goldreich2001}Goldreich, O.: Foundations of Cryptography: Basic Tools. Cambridge University Press, Cambridge (2001)
\bibitem{Gagliardoni2016}Gagliardoni, T., H邦lsing, A., Schaffner, C.: Semantic security and indistinguishability in the quantum world. In: M. Robshaw and J. Katz (Eds.) CRYPTO 2016 Part III. LNCS, vol. 9816, pp. 60-89 (2016)
\bibitem{Shannon1949}Shannon, C.: Communication theory of secrecy systems. Bell Syst. Tech. J. 28(4), 656-715 (1949)
\bibitem{Boneh2013}Boneh, D., Zhandry, M.: Quantum-secure message authentication codes. In: Johansson, T., Nguyen, P. (eds.) EUROCRYPT 2013. LNCS, vol. 7881, pp. 593-609. Springer, Heidelberg (2013)
\bibitem{Broadbent2015}Broadbent, A., Jeffery, S.: Quantum homomorphic encryption for circuits of low T-gate complexity. CRYPTO 2015.
\bibitem{Goldreich1986}Goldreich, O., Goldwasser, S., Micali, S.: How to Construct Random Functions. Journal of the ACM, 33(4):792-807, 1986.
\end{thebibliography}
\end{document}